# Manipulate elastic waves with conventional isotropic materials

Hexuan Gao, Zhihai Xiang[*]

Department of Engineering Mechanics, Tsinghua University, Beijing 100084, China

**Abstract:** Transformation methods have stimulated many interesting applications of manipulating electromagnetic and acoustic waves by using metamaterials, such as super-lens imaging and cloaking. These successes are mainly due to the form-invariant property of the Maxwell equations and acoustic equations. However, the similar progress in manipulating elastic waves is very slow, because the elastodynamic equations are not form-invariant. Here we show that the expression of the elastodynamic potential energy can almost retain its form after conformal mapping, if the longitudinal wave velocity is much larger than the transverse wave velocity, or if the wavelength can be shortened by converting the waves into surface modes. Based on these findings, it is possible to design and fabricate novel devices with ease to manipulate elastic waves at will. One example presented in this paper is an efficient vibration isolator, which contain a 180-degree wave bender made of conventional rubbers. Compared with a conventional isolator of the same shape, similar static support stiffness and smaller damping ratio, this isolator can further reduce wave transmissions by up to 39.9dB in the range of 483 to 1800 Hz in the experiment.

## I. INTRODUCTION

A popular paradigm to manipulate waves is using transformation method, which was originated in manipulating electromagnetic wave [1,2], and has been extended to other types of waves, such as acoustic wave [3,4], surface water wave [5], and even matter wave [6]. However, in contrast to the great success of transformation optics [7-9] and transformation acoustics [10-12], the progress of transformation elastics is rather slow. This is because the classical elastodynamic equations are generally not form-invariant [13], which is the essential prerequisite of transformation methods [1]. To circumvent this difficulty, people have tried to design elastic metamaterials in some special cases, such as the Cosserat type material [14,15], high-frequency approximations [16,17], bending waves in thin plates [18-21], and utilizing the pre-stresses [22-24]. Unfortunately, it is still very difficult to use either of these special solutions to manipulate broadband low-frequency elastic waves of complex modes [25].

[*] Correspondence to: xiangzhihai@tsinghua.edu.cn.

Because of this thorny issue of lack of invariance for elastodynamic equations [9], it is difficult to expect perfect metamaterials to manipulate elastic waves of complex modes. Instead, we can design almost perfect elastic metamaterials based on an idea of minimizing the extra term in the expression of elastic potential energy after transformation. This can be generally achieved by using conformal mapping, requiring the longitudinal wave velocity being much larger than the transverse wave velocity, and converting the wave into surface modes of slow velocities. Since the conformal mapping [2, 17, 26] is adopted in the design, the metamaterial can be easily fabricated with conventional isotropic materials free of local resonance, and has good performance in broad frequency bandwidth. A particular example of the elastic metamaterial designed in this way will be reported here is an elastic wave isolator, which is composed of an aluminum alloy shell and a 180-degree elastic wave bender made of two kinds of conventional rubbers.

This paper is organized as follows. In Section II, we present the theoretical derivation of the effective properties of the elastic metamaterial and an index $\eta$, which measures the extent of form-invariance of potential energy expression after conformal mapping. Based on this theory, we design a vibration isolator in Section III. Numerical simulations are also presented to demonstrate its performances. Then, experimental verifications are detailed in Section IV. Finally, summaries are given in Section V.

**II. THEORETICAL BASIS**

Transformation methods used to design metamaterials [1, 2] usually start from a homogeneous and isotropic material in the virtual space. The geometry of this material in the virtual space is transformed into the desired geometry of a metamaterial in the physical space by using certain coordinate mapping. In this way, a point $x$ in the virtual space is transformed into a point $x' = x'(x)$ in the physical space. At the same time, field variables in the virtual space is also transformed into the physical space with a given gauge. Consequently, by using the change of variables, the governing equations in the virtual space are transformed into the physical space. If the transformed governing equations have the same forms as those in the virtual space, we can easily obtain the effective inhomogeneous material properties of the metamaterial in the physical space by comparing the corresponding terms in these two sets of equations. The selection of the gauge connecting the field variables before and after transformation could be arbitrary [15], and has nothing to do with the form-invariance of the governing equations [24]. However, this gauge is usually chosen as the deformation gradient $\partial x'/\partial x$ [1, 2, 13], so that the deformed grid in physical space shows the wave trajectories [27] and the transformed elasticity tensor is symmetric [15].

Milton et al. has proved that the form-invariance of classic elastodynamic equations are not form-invariant [13]. This is also true if using conformal mappings [28]. All existing work on discussing of form-invariance of elastodynamic equations involve the complex analysis of the gradient of elasticity tensor, because the metamaterial is inhomogeneous. To avoid this complexity, we focus on the form-invariance of the elastic potential energy (integral invariant) instead of the elastodynamic equations (differential invariants) [29] in this paper.

Neglecting the body force, the elastic potential energy in the virtual space of volume $V$ is

$$\Pi = \int_V \frac{1}{2}\varepsilon_{ij}D_{ijkl}\varepsilon_{kl}\mathrm{d}V + \int_V \rho \ddot{u}_i u_i \mathrm{d}V + \int_V \mu \dot{u}_i u_i \mathrm{d}V - \int_{S_T} T_i u_i \mathrm{d}S, \tag{1}$$

where $D_{ijkl} = \lambda \delta_{ij}\delta_{kl} + G(\delta_{ik}\delta_{jl} + \delta_{il}\delta_{jk})$ is the elasticity tensor with Lamé parameters $\lambda$ and $G$; $\delta_{ij}$ is the Kronecker delta; $\varepsilon_{ij}$ is the strain tensor; $u_i$ is the displacement vector; $T_i = \bar{\sigma}_{ij}n_j$ is the given traction vector on boundary $S_T$ with normal vector $n_j$; $\bar{\sigma}_{ij}$ is the stress tensor on boundary $S_T$; $\rho$ is mass density; $\mu$ is the damping factor; and the overhead dot denotes the derivative with respect to time. Here we use the Einstein's notation of index summation. Since conformal mapping will be used in the following, all indices are taken the value of 1 or 2.

We also use the deformation gradient as the gauge to connect the field variables before and after transformation:

$$u_i = u'_j \frac{\partial x'_j}{\partial x_i} \text{ and } \bar{\sigma}_{ij} = J\bar{\sigma}'_{kl}\frac{\partial x_k}{\partial x'_i}\frac{\partial x_l}{\partial x'_j}, \tag{2}$$

where $J = \det(\partial x'_j / \partial x_i)$, and all values with superscript prime are defined in the physical space. In addition, it is well known that the conformal mapping used for the geometrical transformation has the following properties:

$$\frac{\partial x'_k}{\partial x_i}\frac{\partial x'_l}{\partial x_i} = J\delta_{kl} \text{ and } \frac{\partial^2 x'_k}{\partial x_i \partial x_i} = 0. \tag{3}$$

Applying the transformation specified in Eqs. (2) and (3) to Eq. (1) and noticing $\mathrm{d}V = J^{-1}\mathrm{d}V'$, $n_j\mathrm{d}S = J^{-1}(\partial x'_i / \partial x_j)n'_i\mathrm{d}S'$, $T'_i = \bar{\sigma}'_{ij}n'_j$ and $\varepsilon_{ij} = (\partial u_i / \partial x_j + \partial u_j / \partial x_i)/2$, we can obtain the transformed elastic potential energy in the physical space:

$$\Pi' = \int_{V'} \left( \frac{1}{2} J\lambda \varepsilon'_{ii}\varepsilon'_{jj} + JG\varepsilon'_{ij}\varepsilon'_{ij} + Ex \right) dV' + \int_{V'} \rho \ddot{u}'_i u'_i dV' + \int_{V'} \mu \dot{u}'_i u'_i dV' - \int_{S'_T} T'_i u'_i dS', \tag{4}$$

where the extra term is

$$Ex = \frac{G}{J} \left( u'_i u'_j \frac{\partial^2 x'_i}{\partial x_k \partial x_l} \frac{\partial^2 x'_j}{\partial x_k \partial x_l} + 2 u'_i \varepsilon'_{rs} \frac{\partial^2 x'_i}{\partial x_k \partial x_l} \frac{\partial x'_r}{\partial x_k} \frac{\partial x'_s}{\partial x_l} \right). \tag{5}$$

So, the expression of elastic potential energy is not form-invariant, either. This coincides with existing findings [13, 28]. We cannot obtain effective material properties of the metamaterial in the physical space by directly comparing Eq. (4) with Eq. (1), unless the contribution of extra term *Ex* to the strain energy can be neglected. The conditions of ignoring *Ex* can be found by conducting the following dimensional analysis.

Under harmonic excitation of magnitude $a'$ and wavelength $L'$ at frequency $f$, the displacement at location $z' = (x'_1, x'_2)$ can be approximated as $u'_i = a' \text{Re}\{\exp[\mathrm{i}(\kappa' \cdot z' - 2\pi f\, t)]\}$, where $\kappa'$ is the wave number vector of magnitude $\langle \kappa' \rangle \sim 2\pi/L'$. Since $\varepsilon_{ij} = (\partial u_i/\partial x_j + \partial u_j/\partial x_i)/2$, the magnitude of the strain in the physical space is about $\langle \varepsilon'_{ij} \rangle \sim \langle \kappa' \rangle \langle u'_i \rangle$. With all these results and noticing that the velocities of longitudinal wave and transverse wave are $C'_p = \sqrt{(\lambda' + 2G')/\rho'}$ and $C'_s = \sqrt{G'/\rho'}$, respectively, we can calculate the ratio of the magnitude of traditional strain energy terms over the magnitude of the extra term:

$$\eta = \frac{\left\langle \frac{1}{2} J\lambda \varepsilon'_{ii}\varepsilon'_{jj} + JG\varepsilon'_{ij}\varepsilon'_{ij} \right\rangle}{\langle Ex \rangle} = \frac{2\left( \dfrac{J\pi C'_p}{L' C'_s} \right)^2}{\left\langle \dfrac{\partial^2 x'_i}{\partial x_k \partial x_l} \right\rangle \left[ \left\langle \dfrac{\partial^2 x'_j}{\partial x_k \partial x_l} \right\rangle + \dfrac{4\pi}{L'} \left\langle \dfrac{\partial x'_r}{\partial x_k} \dfrac{\partial x'_s}{\partial x_l} \right\rangle \right]}. \tag{6}$$

If $\eta \gg 1$, the potential energy is almost form-invariant:

$$\Pi' \approx \int_{V'} \left( \frac{1}{2} J\lambda \varepsilon'_{ii} \varepsilon'_{jj} + JG \varepsilon'_{ij} \varepsilon'_{ij} \right) dV' + \int_{V'} \rho \ddot{u}'_i u'_i dV' + \int_{V'} \mu \dot{u}'_i u'_i dV' - \int_{S'_T} T'_i u'_i dS'. \tag{7}$$

In this case, the material properties of the metamaterial in the physical space can be obtained by comparing Eq. (7) with Eq. (1) as:

$$\lambda' = J\lambda, \quad G' = JG, \quad \rho' = \rho, \quad \mu' = \mu. \tag{8}$$

## III. REALIZING A VIBRATION ISOLATOR

### A. Effective material properties of the wave bender

With the theory presented in Section II, it is possible to design many interesting elastic metamaterials. Here just gives a simple example, a wave bender, for demonstration.

As Fig. 1 shows, the design starts from a rectangular virtual space filled with homogeneous and isotropic material. This virtual space (described by the point $Z = x_1 + ix_2$ of the complex plane) is transformed into a semi-circular ring shaped physical space (described by the point $z = x'_1 + i x'_2$) with the conformal mapping $z = e^Z$. In this case, we have

$$J = \frac{x_1^2 + x_2^2}{\Delta^2} = \frac{r^2}{\Delta^2}, \tag{9}$$

$$\left\langle \frac{\partial x'_r}{\partial x_k} \frac{\partial x'_s}{\partial x_l} \right\rangle \sim J, \quad \left\langle \frac{\partial^2 x_i}{\partial X_k \partial X_l} \right\rangle \sim \frac{\sqrt{2}r}{\Delta^2}, \tag{10}$$

where $\Delta = 1\,\text{m}$ to make $J$ dimensionless. Consequently, Eq. (6) becomes

$$\eta = \frac{\pi^2 r^2}{L'^2 + 2\sqrt{2}\pi rL'} \left( \frac{C'_p}{C'_s} \right)^2. \tag{11}$$

To ensure a sufficiently large $\eta$, we set $\rho = 1000\,\text{kg/m}^3$, $\lambda = 673.91\,\text{MPa}$ and $G = 0.07\,\text{MPa}$. According to Eq. (8), material properties of this semi-circular ring in the physical space are:

$$\lambda' = \frac{r^2}{\Delta^2}\lambda, \quad G' = \frac{r^2}{\Delta^2}G, \quad \rho' = \rho, \quad \mu' = \mu. \tag{12}$$

Since $C'_p = \sqrt{(\lambda'+2G')/\rho'} = C_p\, r/\Delta$ and $C'_s = \sqrt{G'/\rho'} = C_s\, r/\Delta$, it is easy to verify that a large $\eta$ can be easily achieved. In addition, since $C'_p$ and $C'_s$ are proportional to $r$, elastic waves will bend inside this metamaterial. This wave bending effect only depends on the distribution pattern of $C'_p$ and $C'_s$ regardless of their specific values, i.e., independent of $C_p$ and $C_s$.

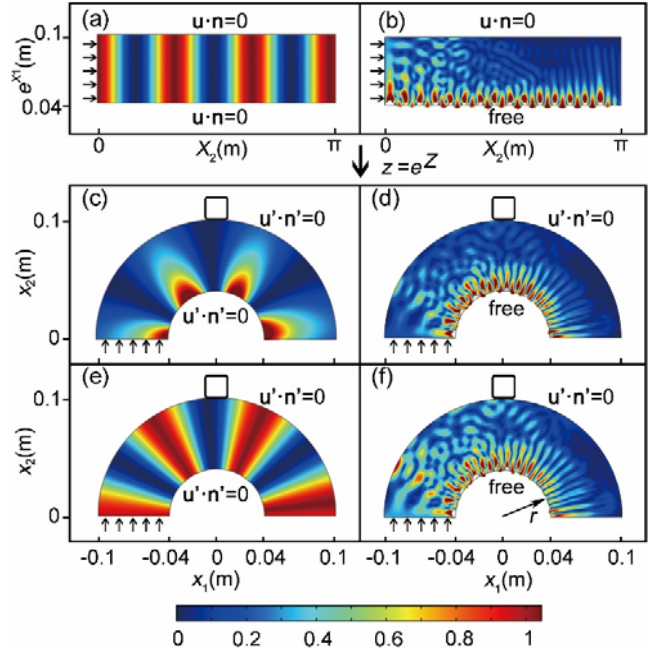

**FIG. 1.** Snap shots of the total displacement (normalized to the amplitude of incident wave) of elastic waves simulated by COMSOL Multiphysics, where $\boldsymbol{u}\cdot\boldsymbol{n}=0$ and $\boldsymbol{u}'\cdot\boldsymbol{n}'=0$ denote the normal constraints on the boundaries in the virtual space and the physical space, respectively. (a) The virtual space with normal constraints on the upper and the lower boundaries ($f = 10000$ Hz). (b) The virtual space with normal constraint only on the upper boundary ($f = 1000$ Hz). (c) The physical space transformed from (a). (d) The physical space transformed from (b). (e) The modified total displacement field $\sqrt{J}\boldsymbol{u}'$ of (c). (f) The modified displacement field $\sqrt{J}\boldsymbol{u}'$ of (d).

With these material properties, we conduct numerical simulations with pure sinusoidal incident longitudinal waves to check the wave bending effect. In Fig. 1(a), since the upper and lower boundaries are normally constrained, only pure longitudinal waves propagate in the virtual space. As Fig. 1(c) shows, the wave does bend in the corresponding physical space with the displacement magnitude inversely proportional to *r*. This coincides with the relation $u_i = u'_j \, \partial x_j / \partial X_i$. Since the incident waves pass along the tangent direction of the outer boundary rather than transmitting through its normal direction, the supported object (denoted by a little square on the top) will not be disturbed. In this way, this wave bender can be used as a novel vibration isolator. In addition, as implied by Eq. (7), the distribution of the modified total displacement $\sqrt{J}\boldsymbol{u}'$ can represent the distribution of strain energy density. This means Fig. 1(d) gives the distribution of strain energy inside the wave bender.

The $\eta$ in Eq. (11) is a crucial index that measures the performance of this wave bender. We can increase $\eta$ by using a material with large ratio of $C'_p / C'_s = \sqrt{\lambda'/G' + 2}$. Since $\lambda' = E'\nu' / [(1+\nu')(1-2\nu')]$ and $G' = E' / [2(1+\nu')]$ ($E'$ is the Young's modulus and $\nu'$ is the Poisson's ratio), the ratio of $C'_p / C'_s = \sqrt{2\nu'/(1-2\nu') + 2}$ can be maximized when $\nu' \to 0.5$. Therefore, it is a natural choice to use rubbers to fabricate this isolator. Eq. (11) also implies that we should decrease the wavelength $L'$. For this purpose, it is preferable to convert the elastic waves into surface modes, e.g. the wavelength of Rayleigh wave is slightly smaller than the transverse wave [30] and much smaller than the longitudinal wave at the same frequency. The wave mode conversion can be achieved by releasing the normal constraint on the boundary, where most waves concentrate near the free surface in Rayleigh mode (see Fig. 1(b), (d) and (f)).

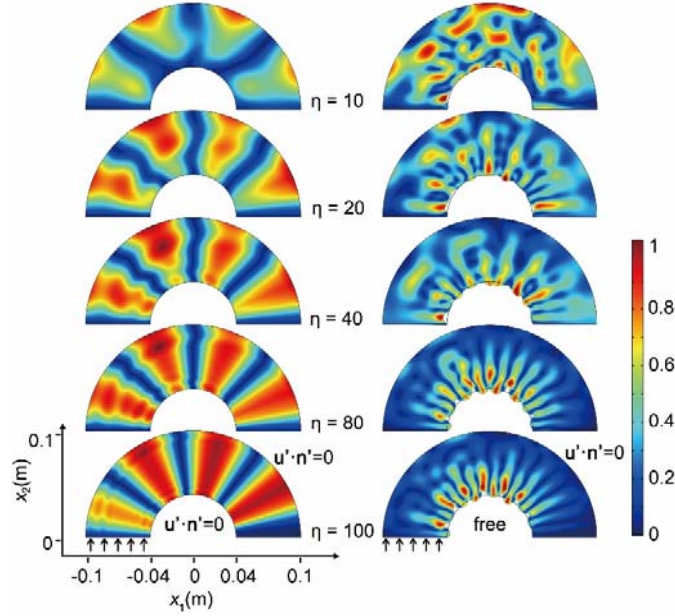

**FIG. 2.** Snap shots of the modified total displacement $\sqrt{J}\boldsymbol{u}'$ (normalized to the amplitude of incident wave) for different $\eta$ simulated by COMSOL Multiphysics.

To check the impact of $\eta$ on the performance of this wave bender, we compare the wave bending effects for different $\eta$ in Fig. 2. In the simulation, $C'_p/C'_s$ is in the range of 9.8 to 31.0, while the wave frequency $f$ is adjusted to keep a constant wavelength of $L'=\pi r/2$ in order to ensure clear comparisons. In this way, $\eta$ is in the range of 10 to 100 according to Eq. (11). The simulation results indicate approximately that good bending effects can be achieved when $\eta > 80$. This means when the extra term is two orders smaller than other terms, the elastic wave equations are nearly form-invariant.

### B. Fabricating the vibration isolator

The effective material properties of this wave bender given in Eq. (12) can be easily realized, because they are isotropic. A simple way is mixing two kinds of conventional rubbers inside a representative cell (see Fig. 3). The material properties of two rubbers adopted here are as follows:

Natural Rubber (NR) $E'_{NR} = 2.61$ MPa, $\rho'_{NR} = 1003$ kg/m³, $C'_{p(NR)} = 1592$ m/s; Silicone Rubber $E'_{SR} = 0.11$ MPa, $\rho'_{SR} = 1092$ kg/m³, $C'_{p(SR)} = 821$ m/s. Since the small difference of densities can be neglected compared with the huge difference of stiffness for these two kinds of rubbers, we can focus on the realizing of $\lambda'$ and $G'$ at different $r$. This can be achieved by adjusting the volume fraction ratio of NR inside a cell at different $r$ (see Fig. 3(c)).

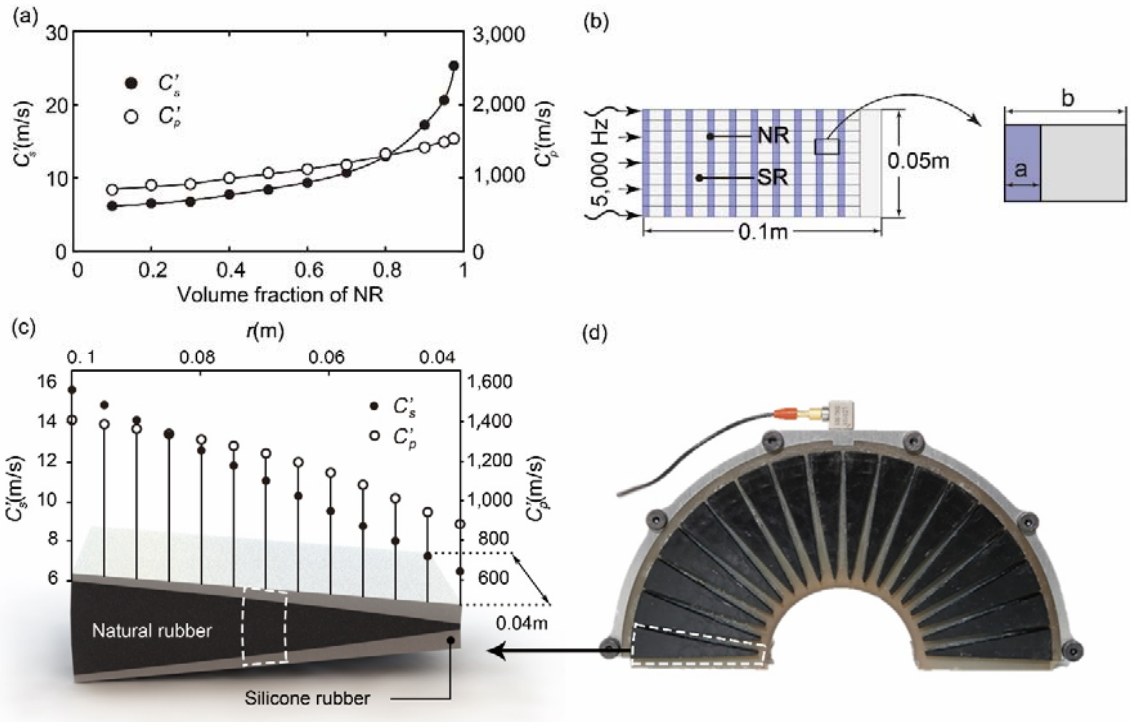

**FIG. 3.** (a) Effective wave velocities obtained from finite element models with different volume fractions of NR, which is solved by COMSOL Multiphysics. (b) Model of the rectangular domain containing 100 cells with the incident wave from left side, and the view of one cell composed of two kinds of rubbers. (c) The fan cell and the distribution of effective wave velocities $C'_p$ and $C'_s$. (d) The vibration isolator with inside wave bender and outside aluminum shell.

The relationships between effective wave velocities and the volume fraction of NR and are obtained by numerical simulations solved by COMSOL Multiphysics. A rectangular domain containing 100 cells is used in the simulation in Fig. 3(b). Each cell is $10 \times 5$ mm and is consisted of two different kinds of rubbers mentioned above. The volume fraction of NR is defined as $a/b$, and it varies from 0.1 to 0.95. For the simulation using longitudinal incident waves, the upper and lower boundaries are normally constrained. While for the simulation using shear waves, tangential displacements are constrained on the upper and lower boundaries. The obtained relationships between effective wave velocities and the volume fraction of NR are plotted in Fig. 3 (a), which indicates that wave velocities (especially the shear velocity) are almost proportional to the volume fraction of NR in the range of 0.0 to 0.8.

Based on Fig. 3 (a), we can design the internal structures of the wave bender with an inner radius of 0.04 m and an outer radius of 0.1 m (see Fig. 3(c)). To facilitate load bearing, the thickness of this bender is set to be 0.04 m. 18 evenly distributed fan cells are in charge of wave bending. Inside each fan cell, there are 13 pair of points on the two interfaces between NR and SR and are evenly distributed along the radius. The circumferential distance of each pair of points determines the volume fraction of NR at position $r$. This distance is determined by the shear velocity curve in Fig. 3 (a) to guarantee that $C'_s$ is strictly proportional to $r$. Applying the curve fitting through these 13 pair of points, obtains the two interfaces between NR and SR. As Fig. 3(c) shows, $C'_p = 1390$m/s and $C'_s = 15.5$m/s on the outer boundary. Referring to Fig. 2, if we require $\eta > 80$ to ensure the bending performance, $L'$ on the outer boundary should be less than 2.74 m according to Eq. (11). This implies that the effective bending of longitudinal waves on the outer boundary approximately starts from the frequency of 507 Hz.

To ensure the integrity of this wave bender, there remains two margins of SR in 5 mm thickness on the inner and outer boundary (see Fig. 3(d)). This does not affect its vibration isolation performance because waves have already been bent inside the wave bender.

The final assembly of the vibration isolator has a 6061 aluminum alloy shell in 5mm thickness on the outer boundary of this wave bender. Since the acoustic impedance of the aluminum shell is much larger than that of the rubber, the shell can be used as normal constraint required on the outer boundary of the wave bender (see Fig. 1(d)). In addition, the silicone oil with a viscosity of 350 cs is used to lubricate the interface between the shell and the rubber, so that the shear force on the interface can be reduced.

## IV. EXPERIMENTS

### A. Basic properties of specimens

To test the performance of the isolator fabricated in Section III (denoted as isolator A), we compare it with another isolator of the same shape (denoted as isolator B). Both isolators have the same aluminum shell, but in contrast to isolator A, isolator B replaces the wave bender (see Fig. 4(a)) with a pure NR of $E = 1.12 \text{ MPa}$ and $\rho = 1072 \text{ kg/m}^3$ (see Fig. 4(b)). A layer of SR is attached at the bottom of the NR in isolator B, so that the measured static support stiffnesses of these two isolators are almost the same (71428 N/m for isolator A and 76923 N/m for isolator B).

Since the damping has great impact on vibration isolation performance, we also measure the damping ratio by using the free vibration test. As Fig. 4(a) and (b) show, low frequency pulses indicated by P are applied on the top of the wave bender and the pure NR with a rubber hammer. The damping ratio $\varsigma$ and the resonance frequency $f_n$ can be extracted from the signals in time domain (see Fig. 4(d)) as $f_{n(A)} = 76 \text{ Hz}$, $f_{n(B)} = 85 \text{ Hz}$, $\varsigma_A = 0.1039$ and $\varsigma_B = 0.1013$. The conclusion of $\varsigma_B < \varsigma_A$ can also be confirmed in Fig. 4(e), because the peak at $f_{n(A)}$ is lower than that at $f_{n(B)}$. The result of free vibration test for the aluminum shell is also presented in Fig. 4 (e), in which we can find two dominant $f_n$ at 162 Hz and 934 Hz, which coincide with the first and the third $f_n$ simulated by COMSOL Multiphysics (Fig. 4 (f) and Fig. 4 (h)).

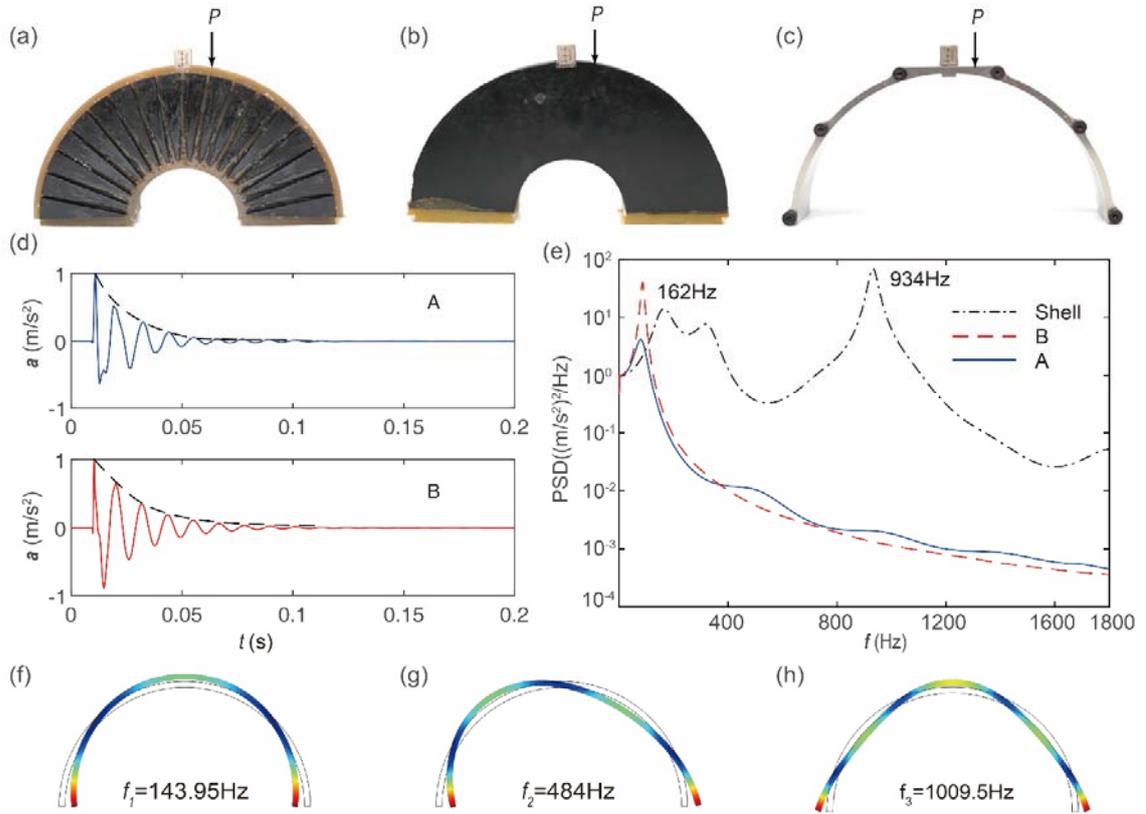

**FIG. 4.** (a) Free vibration test of the wave bender. (b) Free vibration test of the pure rubber. (c) Free vibration test of the aluminum shell. (d) Acceleration signals on the top of the wave bender (A) and the pure rubber (B), normalized to the maximum amplitude of the pulse P. (e) Power spectrum density of the acceleration signals. (f), (g), (h) First three vibration modes of the aluminum shell solved by the COMSOL Multiphysics.

## B. Simulation results

Before the real testing, the performances of these two isolators are simulated by using the COMSOL Multiphysics. Since it is very difficult to accurately determine the damping inside the complex assembly of these isolators, we use the same damping ratio of 0.1 in the simulations. "The Thin Elastic Layer boundary condition" in COMSOL Multiphysics with normal stiffness of $k_n = 1 \times 10^{13}$ N/m$^2$ and tangent stiffness of $k_t = 0$ N/m$^2$ are used on the interface between the shell and rubber.

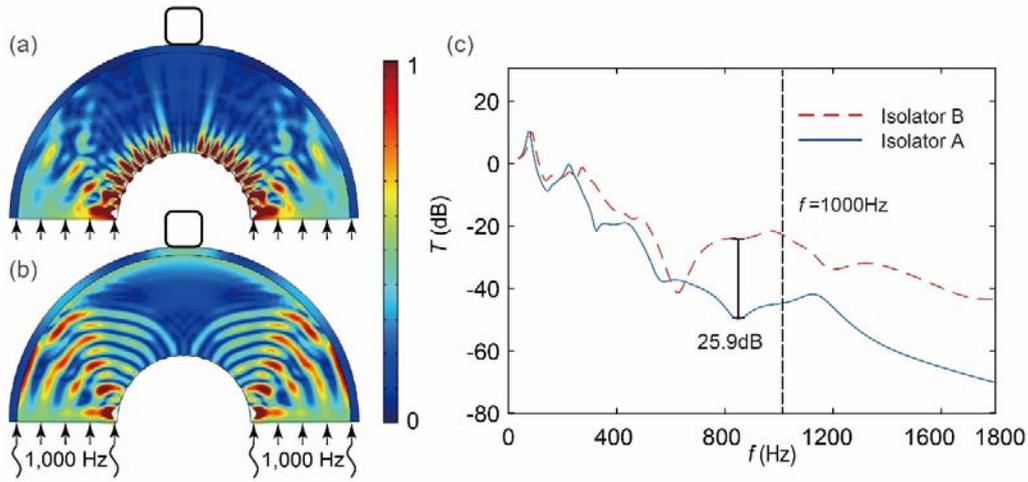

**FIG. 5.** (a) Snap shot of the total displacement field (normalized to the amplitude of incident wave) of isolator A at 1000Hz. (b) Snap shot of the total displacement field (normalized to the amplitude of incident wave) of isolator B at 1000Hz. (c) Transfer functions of isolator A and isolator B.

As Fig. 5(a) shows, most of the incident waves are converted into the Rayleigh mode and bent along the inner boundary of isolator A, so that there is just little disturbance on the aluminum shell. While as Fig. 5(b) shows, large fraction of waves directly transmit through the aluminum shell of isolator B, so that it could have poorer isolation performance than that of isolator A. These observations can be further confirmed in Fig. 5(c), where the transfer function of isolator A has the maximum reduction of 25.9 dB at 853 Hz. The transfer function is defined as $T(f) = 20\log_{10}\left[u_{top}(f)/u_{input}(f)\right]$, where $u_{input}(f)$ and $u_{top}(f)$ are the amplitudes of incident waves and the top of isolators at frequency $f$, respectively.

### C. Experimental results

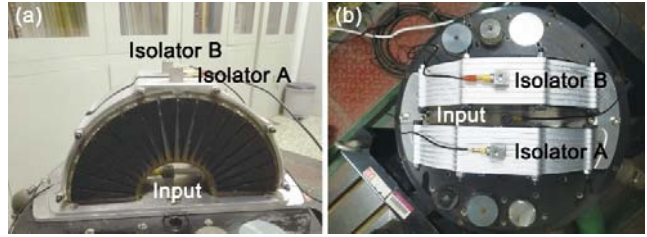

**FIG. 6.** The setup of shaking table test. (a) Front view. (b) Top view.

As Fig. 6 shows, shaking table test with sweep frequency input signals is firstly conducted to test the real performances of these two isolators. These isolators are placed closely on an aluminum base mounted on the shaking table. An accelerator is fixed in the center of the aluminum base to measure the input signal generated by the shaking table. Other two identical accelerators are fixed on the top of isolator A and isolator B, respectively. The recorded accelerations are plotted in Fig. 7(a), which shows the obvious reduction of amplitude by using isolator A. More detailed comparisons are given in Fig. 7(c). The transmissibility is defined as $H(f) = 10\log_{10}\left[X_T(f)/X_I(f)\right]$, where $X_T$ denotes the power spectrum density of the acceleration measured on the top of isolator A or B, and $X_I$ denotes the power spectrum density of the input acceleration measured on the aluminum plate. Since isolator A has little bit larger damping ratio than that of isolator B as we measured in Fig. 4, its transmissibility around resonance is smaller than that isolator B. According to the traditional passive isolation theory [31], isolator A should be less effective than isolator B at frequencies above the resonance. However, as Fig.

7(c) shows, after $f > 483$Hz isolator A surpasses isolator B and achieves its best performance at $f = 833$Hz with 39.9dB further reduction of transmissibility. These observations consistent with the estimation of start frequency of 507Hz given in Section II.B and the best performance frequency of 853 Hz given in Fig. 5 (c). In addition, the bending effect disappears at high frequencies. This is because compared with the circumferential size of a cell, the wavelengths are not large enough to ensure the effective wave velocities.

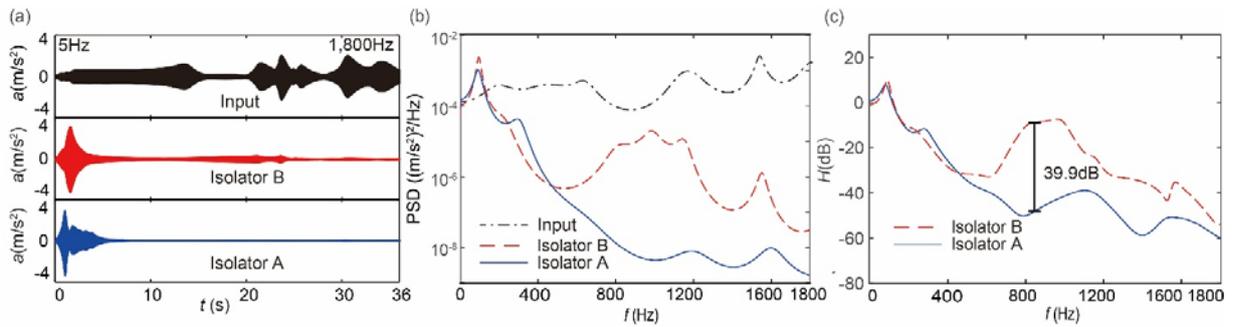

**FIG. 7.** Results of sweep frequency test. (a) Accelerations measured on the aluminum base (input), the top of isolator B and the top of isolator A. (b) The power spectrum density of the accelerations (c) The transmissibility of isolator A and isolator B.

Besides the sweep frequency test, the tap test is also conducted for comparison. In the tap test, the positions of two isolators and three accelerators are the same as those of the sweep frequency test (see Fig. 6). While input signal is generated by tapping on the aluminum base using a steel hammer. The recorded signals are plotted in Fig. 8(a), which confirms that isolator A super performs isolator B. Detailed comparison in frequency domain is presented in Fig. 8(c). Similar to the sweep frequency test, after $f > 387$Hz isolator A surpasses isolator B and achieves its best performance at $f = 910$Hz with 44.1dB further reduction of transmissibility. However, we cannot find peaks at resonant frequencies, because the input signal due to tapping is mainly at high frequencies, so that the power around $f_n$ is not high enough to excite the resonance of the isolators.

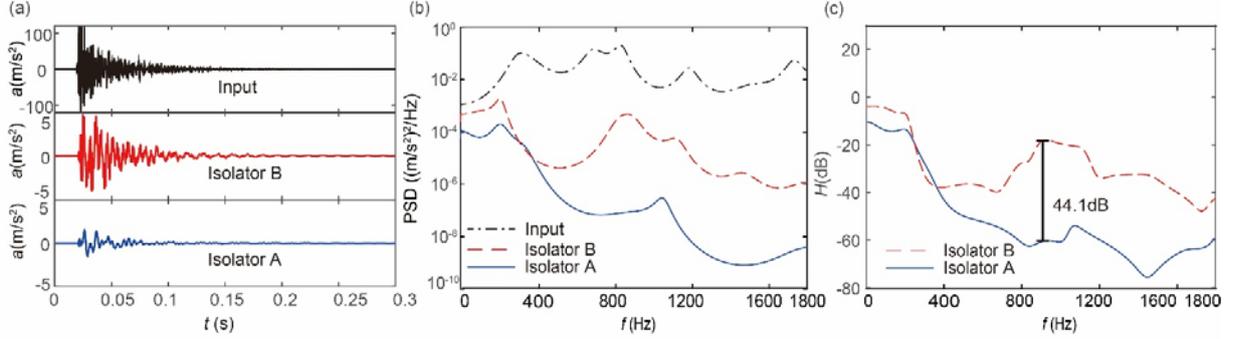

**FIG. 8.** Results of tap test. (a) Accelerations measured on the aluminum base (input), the top of isolator B and the top of isolator A. (b) The power spectrum density of the accelerations (c) The transmissibility of isolator A and isolator B.

## V. SUMMARY

In this paper, we propose a general method to design elastic metamaterials based on the conformal transformation. The great challenge of form-invariance is solved by using the material whose longitudinal wave velocity is much larger than the transverse wave velocity. In addition, the performance of the resultant elastic metamaterial can be further improved by introducing interfaces to convert elastic waves into the modes with short wavelengths. The validity of this design method is demonstrated by an elastic wave bender acting as a vibration isolator. Although this is a simple example, it shows superior performance at broad frequency bandwidth and even breaks the limit on damping ratios required by the classic passive vibration theory. Thus, we can envisage that other interesting implementations of elastic metamaterials can be achieved in future by using the method proposed in this paper.

**Acknowledgments:** This work was supported by the National Natural Science Foundation of China with the grant number 11672144. The assistance of B. Wang and Q. H. Lu for the shaking table test is greatly appreciated.